# Axion Generation in a Three-Dimensional Optical Trap

**Chunyu Zhang** [1]✉, Xinran Fang[1], Lichen Peng[2], Xuri Yao[2]✉, Xiaoying Tang[1]✉

The axion is a theoretical particle that could resolve multiple fundamental problems, most notably the strong Charge-conjugation-parity-symmetry (CP) problem in quantum chromodynamics and the nature of dark matter. To date, however, the axion has never been detected in any free-space experiment. In this work, we designed and constructed a laser-based system that generates a three-dimensional, closed trapping potential field with a null central region. Owing to its spindle-like geometry, we term this configuration an optical spindle trap (OST). Along the propagation axis, the photon population evolves in a distinct manner from the left to the right terminus of the trap: it first diminishes and then recovers to its baseline value. This behavior is analogous to the photon–axion–photon conversion process sought in light-shining-through-wall experiments (LSW)[1-3], in which a measured photon deficit would constitute evidence for axion conversion. The photon population was monitored with a single-photon counter (SPC) operated well below its saturation threshold, and the observed behavior was corroborated by charge-coupled device (CCD) imaging at extremely low optical powers, thereby excluding detector artefacts as the origin of the photon deficit. Under the constraint of energy conservation, the missing photons are attributed to conversion into axions that remain undetectable by both the SPC and the CCD. The underlying conversion mechanism is ascribed to spin-coupled axion–photon interactions. This tens-of-millimeter-scale optical spindle trap thus provides a viable free-space axion source, generated by a table-top laser, for the study of the strong CP problem and axion-like dark matter candidates.

[1]School of Medical Science and Engineering, Beijing Institute of Technology, Beijing 100081, China. [2] School of Physics, Beijing Institute of Technology, Beijing 100081, China.

[*]Correspondence and requests for materials should be addressed to Chunyu Zhang. E-mail: zzccyy1977@bit.edu.cn

The existence of dark matter and dark energy is firmly supported by a wealth of astronomical observations[4–7]. Numerous theoretical frameworks have been proposed to account for these phenomena, including weakly interacting massive particles (WIMPs), axions and axion-like particles, light bosons, and neutrinos[8–12]; however, to date, direct and compelling detection of these dark matter candidates remains elusive. Among these, the axion stands out as a prominent candidate whose characteristics more closely resemble those of conventional electromagnetic waves. The axion originates from a proposed resolution to the strong CP problem in quantum chromodynamics (QCD), first introduced by R. D. Peccei and H. R. Quinn in 1977[13,14]. Subsequently, Weinberg and Wilczek independently demonstrated that a natural consequence of the spontaneous breaking of this Peccei–Quinn (PQ) symmetry is the existence of an extremely light, hypothetical particle—the axion[15,16]. Despite extensive efforts, the search for axions continues in both astrophysical contexts and terrestrial laboratory experiments[17,18]. Should the axion exist with a sufficiently small mass, it would not only constitute a well-motivated candidate for cold dark matter but would also elegantly resolve the strong CP problem in QCD, thereby bearing profound implications for both particle physics and cosmology. Even if axions do not constitute the entirety of dark matter, their existence remains a distinct theoretical possibility[19]. In 2025, the observation of a dynamical axion quasiparticle in the material $MnBi_2Te_4$ was reported[20,21]. Furthermore, comprehensive reviews in Science and Reviews of Modern Physics, along with the U.S. Snowmass white papers (e.g., Snowmass 86, 13 (M3 1)), have consistently identified strongly correlated quantum fields as among the most promising avenues for the generation and detection of axions[22–24]. Notably, the singularity within the laser field employed in this experiment provides precisely the requisite conditions for realizing such strongly interacting quantum fields. This context aligns with the objectives outlined in the Snowmass 2021 process concerning eV-scale dark matter and the development of single-photon detectors with low dark counts[22].

Here, we report the observation of axion generation within a free-space laser field.

In standard LSW[1-3] configurations, axions may be sourced by matter density alone and subsequently detected via their characteristic coupling to electromagnetism. This process entails the conversion of a minority photon population into axions, which then traverse an optical barrier before being reconverted into photons under an applied magnetic field, manifesting as a distinct photon deficit followed by recovery. In the present experiment, we observe precisely this signature within an optical spindle trap. Under conventional conditions devoid of frequency doubling, absorption, or reflection losses, photon number is expected to be strictly conserved. Crucially, the photon deficit observed in this work cannot be attributed to mere optical interference effects responsible for trap formation; instead, it constitutes evidence for axion production, wherein the diminished photons are converted into axions undetectable by the single-photon counter. This interpretation is consistent with theoretical predictions that axions are considerably lighter than WIMPs and interact with ordinary matter exclusively through electromagnetic (and gravitational) couplings. Whereas direct axion detection experiments typically rely on magnetic fields to induce photon reconversion, the quantum nature of axions becomes particularly pronounced owing to their exceedingly small mass. In this context, LSW experiments are generally more challenging than direct dark matter searches, as both axion production and detection entail additional powers of small coupling constants. In the scenario explored here, axions can be sourced by matter density alone and subsequently probed via their standard couplings to electromagnetism or spin degrees of freedom. Various experimental schemes have been proposed along these lines, including those employing spin probes to detect axions generated by laboratory sources[25] or by the Earth itself[26-30]. The present work demonstrates that the singular optical field structure of a millimeter-scale spindle trap offers a viable, free-space platform for accessing these elusive axion couplings.

## Experimental setup

A apparatus (Extended Data Fig. 1) enables

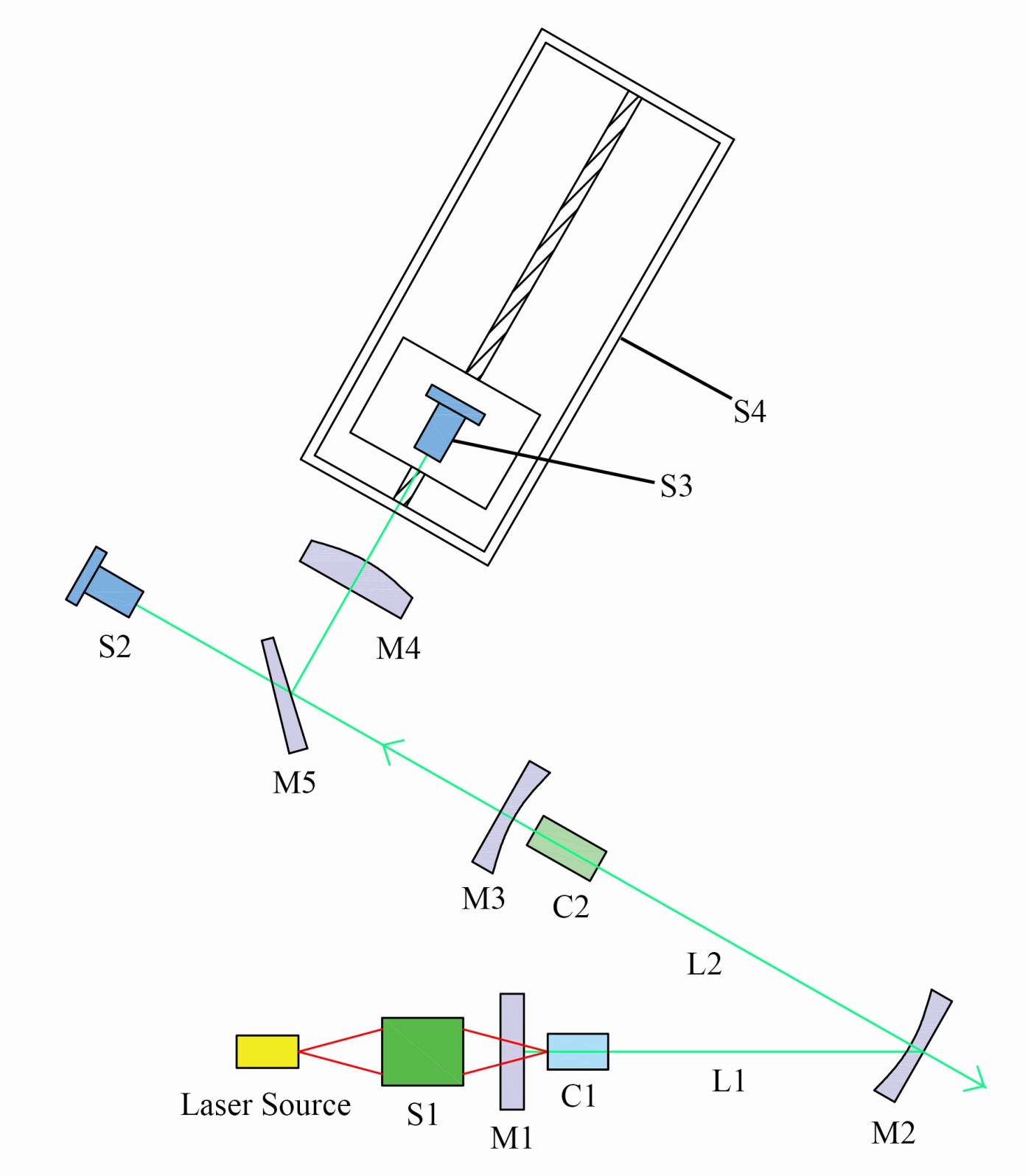

Fig.1| Laser diode 808 nm, S1: coupling system; M1, M2, M3: cavity mirrors coating High refection at 1064 nm, M2, M3 coating AR 532 nm, M1 or M3 defect coating or spot (circle micro-machined) at 30 um; A1: Nd: YVO4; A2: LBO, nonlinear crystal; M4: convex lens; M5: optical wedge; S2, S3: single photon counter (Hamamatsu H10682); S4: translation.

Here we outline the generation of a three-dimensional confined OST. The procedure begins with a 1064 nm Gaussian oscillation. Intracavity phase modulation is then employed to produce a $TEM_{01}$ vortex beam, after which frequency doubling in a nonlinear LBO crystal yields green light in the $TEM_{10}$ mode—a process that we previously examined both theoretically and experimentally in the context of 586 nm yellow emission[31]. Subsequently, the $TEM_{10}$ mode is shaped by a single lens: by tuning the lens position and focal length, a three-dimensional hollow energy trap forms near the focal plane. A sequence of 850 transverse beam profiles was recorded, and the resulting stack was assembled in ImageJ to generate the composite image presented in Fig. 3. The beam produced in this manner closely resembles an optical bottle beam. Conventional generation of such beams typically relies on extra-cavity phase modulation using spatial light modulators (SLMs), holographic phase plates, or similar techniques[32–37]. These approaches depend heavily on the fabrication fidelity and performance of the optical elements, and the intensity at the beam center often cannot be brought fully to zero. An alternative scheme involves two-beam interference, which however entails complex architectures, constrained mechanical tunability, limited stability and low coherence, and is frequently accompanied by spurious diffraction orders. By contrast, the methodology reported here is based on intracavity laser field manipulation in conjunction with a nonlinear optical process. Through two sequential stages—intracavity resonant mode selection followed by nonlinear mode selection—we obtain a single transverse mode, devoid of any residual fundamental or higher-order components. The resulting single-beam output is markedly stable, exhibits enhanced photon degeneracy, improved optical coherence, and better quantum characteristics.

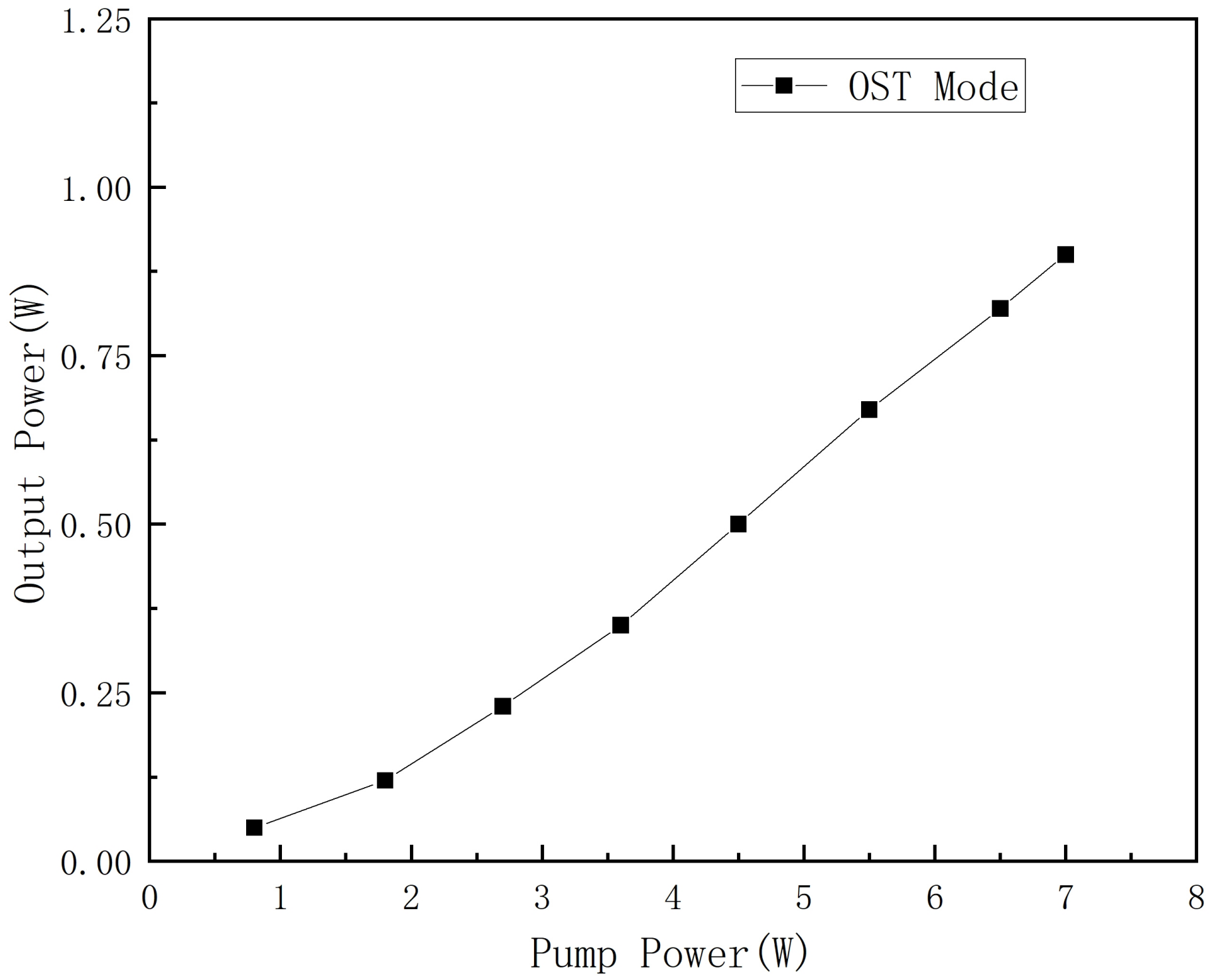

Fig. 2 | Laser power input–output curve. At a pump power of nearly 7 W, the power in the shuttle-shaped optical trap reaches 1 W.

We acquired several hundred images of the beam cross-section using a CCD camera. Image-processing software was used to stack these images and reconstruct the three-dimensional structure, which is consistent with direct visual observation, as shown in Fig. 3b. The optical trap obtained in our experiment differs markedly from the optical bottom beam reported previously[33-35]. In the optical bottle beam, the central photons undergo an abrupt circulation to the outer ring, whereas no such phenomenon is observed here. Instead, the central intensity gradually decreases to near zero over a couple of millimetres and then symmetrically recovers to the original profile. At present, the maximum output power of the shuttle-shaped optical trap beam is 600 mW, with the corresponding power input–output curve presented in Fig. 2. If the third step is omitted, the resulting Gaussian beam can reach a power of 800 mW, indicating that these three steps cause a loss of approximately 30%. Currently, the 532 nm Gaussian beam can generate powers of several tens of watts, with further replacement of components, output powers exceeding 10 W can be readily achieved.

## Discussion

In Fig. 3c, the position and parameters of M4 were adjusted so that the laser approached a collimated state. As the central intensity gradually decreased to zero and then recovered, the energy of the outer ring first increased and then decreased. To resolve the internal structure more clearly, Fig. 3c was compressed horizontally and stretched vertically to generate Fig. 3d. Under these conditions, photons in region 1 gradually decreased, transitioned through region 2 via a specific modality, and were transferred to region 3, where the photon count progressively increased. Subsequently, energy was transferred from region 3 to region 5 through region 4 via another specific modality. The particles responsible for these specific modalities remain undetectable at present.

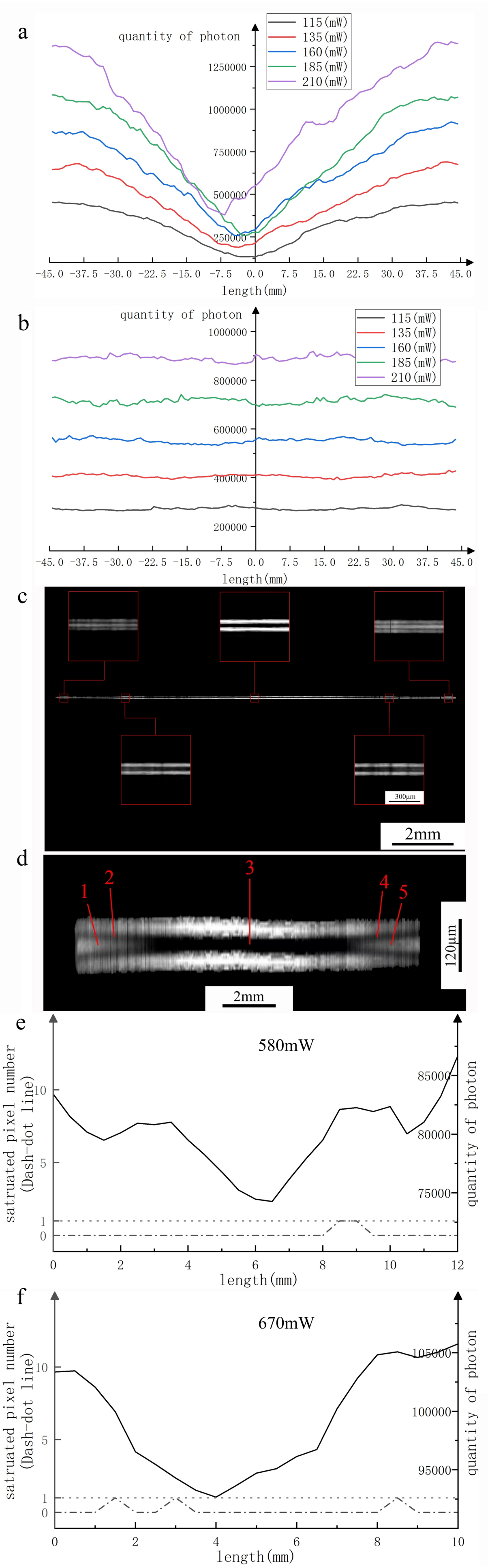

a
quantity of photon
115(mW)
135(mW)
160(mW)
185(mW)
210(mW)
length(mm)
b
quantity of photon
length(mm)
c
300μm
2mm
d
1
2
3
4
5
120μm
2mm
e
580mW
satruated pixel number (Dash-dot line)
quantity of photon
length(mm)
f
670mW
satruated pixel number (Dash-dot line)
quantity of photon
length(mm)

**Fig.3**| Beam cross-sectional profiles and photon-number distributions.
**a,** The photon-number curves measured at different optical powers (S3 data denoised via S2 data) display a symmetric profile, featuring an initial decline followed by a subsequent rise.
**b,** Original S2 data, showing no intensity dip.
**c,** The central beam profile reconstructed from a stack of cross-sectional images acquired by a CCD camera, with a vertical cross-section taken along the propagation direction through the mid-plane. Magnified views of selected local regions along the beam are provided.
**d,** The image in c further compressed in the transverse direction and stretched longitudinally, revealing a refined internal structure: the central beam progressively weakens, narrows, and eventually vanishes, followed by a symmetric recovery in intensity and width; concomitantly, the peripheral beam first intensifies and then attenuates.
**e** and **f,** Photon-number distribution recorded on the CCD at different power of lasers, also displaying a dip-and-recovery pattern. All photon-number measurements—both from the single-photon counter and the CCD—were performed well below the saturation threshold to preclude any instrument-induced photon-number reduction due to saturation. The lower dash-dotted lines in e and f indicate the saturation pixel count of the CCD. In a, c, e and f, all profiles are symmetric with respect to the beam center.

Two interpretations are proposed based on the wave–particle duality of light. First, considering the wave nature — notably interference — the central beam and the outer ring interfere at and around the optical trap center, leading to photon redistribution. Interference coupling near point 6 results in energy reallocation, increasing the energy in region 3. In our experimental observations, no energy flow bypassing the singularity at the trap center was observed; a dark region always separated the central part from the outer ring. Using the 1064 nm annular beam leaked from cavity mirror M3 under identical focusing conditions, we confirmed that the beam remained annular before and after the focal plane. Photons in the outer ring cannot reach the center, and the dark region along the optical axis — where the central beam resides near the axis — should persist. This is consistent with the conventional understanding that higher-order modes passing through an ideal lens maintain their mode type, merely undergoing isotropic scaling. That is, a residual on-axis bright spot from the central beam would be expected. However, experimentally, the center remained dark, exhibiting a singularity. The spin phase matching condition here plays a role analogous to a strong magnetic field in the photon–axion interconversion. The singularity arises strictly at the optical trap center, and instead of isotropic scaling near the focal plane, a strong interaction of light quanta occurs. The twisted spin phase gives rise to the singularity, where the θ-term of the strong CP problem in QCD emerges — leading to axion production. The axion acts as a carrier of energy transfer, such that most photons from the central bright spot are transferred to the outer ring. The missing fraction, representing an intermediate state, carries energy in the form of axions. Consistent with the notion that axions are sourced by regions with aligned electric and magnetic fields, in our laser system this region is extremely small, producing a low number of axions with correspondingly limited energy. Direct measurement remains a viable approach, yet to date no direct evidence of axions has been observed anywhere in air or space—most likely due to insufficient sensitivity[38], however, Considerable theoretical attention has been devoted to the possibilities and mechanisms of photon–axion coupling[39-42]. Here we adopt an indirect method based on total energy conservation. By measuring the photon numbers in the transverse cross-section at the center and both sides of an optical trap, we find that any minute decrease in the total photon count would indicate the presence of axions. Our experiments confirm this expectation: the photon number decreases before the singularity and returns to its original value after the singularity, as shown in Fig. 3a. The reduction in photon number accounts for approximately 5% of the total. Whether the crossing of optical paths due to lenses causes interference, or the reduction in beam volume induces strong interactions of light quanta, or a combination of both effects facilitates axion generation, the singularity distorts both temporal and spatial energy distribution. Consequently, this laser system offers a versatile experimental tool for the broader physics community.

Second, from the perspective of the particle nature of light, energy originates in a resonant cavity and propagates at the speed of light along the beam. After the convex lens, the central spot carries approximately half of the total energy. Under specific conditions, photons in the central beam become coupled in a certain manner. A dark region always exists between the central beam and the outer ring. Energy comprising half of the beam’s total power is transferred from region 1 to region 3 across this dark region. In other words, energy from the central part is transferred to the outer ring in some form. Upon a change in spin–coupling conditions, energy returns from region 3 to the center across the dark region. This process closely resembles the core principle of the wall-separation experiment used in axion searches. Regrettably, the LSW experiment has never observed the reappearance of photons. By contrast, our experiment captures this process: the photon number first decreases and then returns to its initial value. The vanished photons must have been converted into some particle form undetectable by conventional detectors, which we identify as axions or axion-like particles—the very particles scientists have long sought. Notably, region 3 exhibits a higher abundance of axions and axion-like particles. This discovery provides an opportunity for the search for dark matter and dark energy.

The laser employed in our experiment possesses excellent quantum properties. Annular beams have already been proven beneficial for quantum communication and quantum computing[43,44]. Our beam builds on the annular beam but is generated through nonlinear mode selection, endowing it with stronger quantum correlations. This aligns with the axion theory proposed to resolve the strong CP problem in QCD. Researchers working on the strong CP problem argue that axions have favorable quantum properties and can interconvert with photons—a principle underpinning most current experiments. Laboratories in the United States, Germany, and elsewhere have long used intense lasers to produce photons and attempted to convert photons into axions via external strong magnetic fields, then observe the reverse conversion[1-3]. Despite decades of effort, no such process has been observed, perhaps due to insufficient magnetic field strength or failure to satisfy the conditions for photon–axion interconversion. In our experiment, without any external strong magnetic field, the photo spin–coupling process—where the spiral phase of the outer ring interacts with the Gaussian planar wave phase at the center—facilitates spin–coupling. Should axion-like dark matter or dark energy be generated in this process, our findings would represent a paradigm shift in the search for these elusive components of the universe.

We attribute the observed photon number reduction not to the single-photon counter (SPC) itself. Specifically: (1) In typical SPC operation, photon losses usually occur when the detected photon count increases, whereas here we first observe a decrease. This reduction is unlikely to result from detector saturation or dead time effects. (2) By attenuating the input light, the photon flux incident on the SPC remained well below the saturation threshold ($5 \times 10^6$ counts). The maximum measured counts ranged from $2 \times 10^5$ to $8 \times 10^5$, and five datasets acquired under different light intensities consistently showed the same decreasing trend. (3) The dead time of the SPC is 20 ns, while our measurements were taken at intervals of 2 s. (4) The same reduction-then-increase trend was reproduced on different detectors (CCD). Additionally, at lower excitation powers, we quantified the photon counts collected by the CCD (Fig. 3 **e** and **f**). As indicated by the dashed line in the figure, at most one saturated pixel appears randomly, and this saturation does not occur at the descending part of the curve; therefore, it cannot account for the observed decline. Collectively, these results confirm that the decrease in photon number is not due to detector artefacts.

**Summary**

Conventionally, interference leads to a spatial redistribution of energy—or equivalently, of photon number—rather than any net reduction, the experimental results reported here suggest the occurrence of a photon–axion–photon conversion process at the tens-of-millimeters scale. Experimentally, we have realized a three-dimensionally confined potential well of energy within a single laser beam using an intracavity modulation-based method. Within this optical field, internal energy redistribution takes place, accompanied by a photon–axion–photon conversion process. The underlying mechanism is spin coupling, which, inside the optical trap, provides a means of generating axions or axion-like particles (ALPs) in free space, thereby offering a new experimental approach for studying the strong CP problem and dark matter candidates.

Astronomically, it is thought that during the Big Bang or the subsequent expansion of the Universe—alongside the formation of galaxies and other celestial bodies—dark matter and dark energy emerged with inhomogeneous density distributions.

Such inhomogeneities, particularly under the influence of gravitational lensing, may facilitate the proposed conversion process. In our experiment, the photon density between the central region and the outer ring of the optical trap is distinctly nonuniform, yet these regions remain strongly correlated through interactions. Under the effect of lensing, the spatial distribution of energy is altered, analogous to astrophysical structure formation. We therefore hypothesize that in certain optically interfering or quantum-strongly-interacting dark regions, axions or ALPs are produced; upon phase mismatch, they convert back into photons. However, owing to their extremely low mass, the miniscule spatial volume in which they are generated, and the associated low energy, conventional power meters are unable to detect the energy difference. Here, we provide experimental evidence for the existence of axion dark matter by directly measuring the number of laser photons.

**Method**

The pump light source is a semiconductor laser of FOCUSLIGHT (FL-DLS02-FCSB04-I-30-808-10), with a maximum output power of 30W, a laser spot diameter of 200 μm and an output wavelength of 808 nm. The parameter of laser output coupling mirror S1 is 1: 0.8. In the laser resonator, the curvature radius of the concave mirror M2 is 200 mm, and the curvature radius of the concave mirror M3 is 100 mm. M1 is a plane mirror with high reflectivity (HR) at 1064 nm; M2 and M3 are concave mirrors featuring HR coatings at 1064 nm and high transmission (HT) at 532 nm. M3 incorporates a microfabricated circular aperture (pinhole) with a diameter of 30 μm. The distances between S1 and M1 are 40 mm, M1 and C1 are 10 mm, C1 and M2 are 160 mm, M2 and C2 are 90 mm, C2 and M3 are 70 mm. C1 is an Nd:YVO4 crystal (3 × 3 × 10 mm) cooled by a water chiller (LX-150, 150 W, 20 °C), while C2 is an LBO crystal (2 × 2 × 12 mm) regulated by a TEC module to maintain its optimal operating temperature of 30 °C. A convex lens with a 100 mm focal length is used as M4. The distance between M3 and M4 is set to 250 mm, resulting in a hollow spot size suitable for CCD observation or single photon counts. The single-photon counter uses the Hamamatsu H10682. The CCD sensor features a resolution of 720 × 576 pixels with a physical pixel size of 5 × 6.25 μm, enabling a measurement accuracy of 6.25 μm. The digital grayscale values output by a CCD are directly proportional to the number of incident photons within a certain linear range. We can estimate the number of incident photons based on these output grayscale values. The model of the S4 stepper motor is 42BYG250C-DH1.

The curve in Fig. 3a is obtained by denoising the S3 data using S2. Specifically, the operation S3 − 2.5 × S2 is applied to eliminate the influence of the reference beam (S2) on the signal beam (S3). Under this coefficient, the dip in the curve is better aligned. The S2 data reflect the laser energy fluctuations in real time.

**DISCLOSURE STATEMENT**

The authors are not aware of any affiliations, memberships, funding, or financial holdings that might be perceived as affecting the objectivity of this review.

**ACKNOWLEDGMENTS**

We would like to sincerely thank Bing Xu, Pengcheng Mao from the Analysis & Testing Center (Beijing Institute of Technology) for providing the necessary space and facilities.

**Author contribution**

C.Z. conceived and designed the study. C.Z. and X.F. constructed the experimental setup and performed the experiments. C.Z. wrote the manuscript. X.F. helped with manuscript revision. C.Z. oversaw and directed the whole project. P.L. participated in some experiments. X.Y. provided the single photon counters and processed the CCD data. X.Y. and X.T. had a meaningful discussion on the experimental content.